\documentclass[twocolumn,showpacs,preprintnumbers,amsmath,amssymb,prb,superscriptaddress]{revtex4}

\usepackage{graphicx}% Include figure files
\usepackage{dcolumn}% Align table columns on decimal point
\usepackage{bm}% bold math

\begin{document}

\preprint{InAs Mn draft6-arXiv}

\title{Influence of surface-related strain and electric field on acceptor wave functions in Zincblende semiconductors}

\author{S. Loth}
 \surname{Loth}
\author{M. Wenderoth}
 \email{wendero@ph4.physik.uni-goettingen.de}
\author{R. G. Ulbrich}
\affiliation{IV.~Physikalisches Institut der Georg-August-Universit\"at
G\"ottingen, Friedrich-Hund-Platz.~1, 37077 G\"ottingen, Germany}%

\date{\today}% It is always \today, today,
             %  but any date may be explicitly specified

\begin{abstract}
The spatial distribution of the local density of states (LDOS) at Mn acceptors near the (110) surface of p-doped InAs is investigated by Scanning Tunneling Microscopy (STM). The shapes of the acceptor contrasts for different dopant depths under the surface are analyzed. Acceptors located within the first ten subsurface layers of the semiconductor show a lower symmetry than expected from theoretical predictions of the bulk acceptor wave function. They exhibit a (001) mirror asymmetry. The degree of asymmetry depends on the acceptor atoms' depths. The measured contrasts for acceptors buried below the 10th subsurface layer closely match the theoretically derived shape. Two effects are able to explain the symmetry reduction: the strain field of the surface relaxation and the tip-induced electric field.
\end{abstract}

% PACS, the Physics and Astronomy
                             % Classification Scheme.

%keywords
%Dopants \sep III-V semiconductors \sep Scanning Tunneling Microscopy and Spectroscopy \sep Surface and interfaces

\maketitle

\section{Introduction}
Scanning Tunneling Microscope (STM) studies of the local electronic contrasts induced by shallow and deep acceptors in III-V semiconductors are subject to intense discussions\cite{zhe94,kor01,yak04,mah05,kit05a,lot06p,yak2007,Mar2007b}. The anisotropic contrasts of magnetic acceptors like Mn are of particular interest, because their microscopic coupling to holes and other Mn acceptors influences the macroscopic magnetic properties of the doped semiconductor\cite{San2002, Sat2004, Kit2006}. For acceptors in Zincblende semiconductors, e.g., the III-V compounds, one expects that the observed contrasts reflect the cubic symmetry of the host crystal's band structure (hence c$_{2v}$)\cite{Bal1974, Mon2006}.
However, shallow acceptors show up in STM topographies as triangular contrasts with the dopant atom located in the triangle's tip, clearly breaking the c$_{2v}$ symmetry\cite{zhe94,kor01,lot06j}.
Deep acceptors show an asymmetric bow-tie like shape reminiscent of the bulk symmetry, but nevertheless asymmetric with reference to the (001) mirror plane\cite{ars2003,yak04}.
Tight binding calculations were performed to describe the acceptor state in the bulk crystal\cite{yak04,fla05}. Up to now, the semiconductor surface was not fully included into such calculations because the necessary large slab would exceed today's computing capabilities.
The probability density at the sample surface that originates from the wave function of a subsurface acceptor was extracted from the existing bulk calculation by cutting the calculated 3D probability density at a certain distance from the acceptor atom and artificially adding the decay into the the vacuum\cite{fla05,Mon2006}.
However, an acceptor in the vicinity of the surface will not only differ in electronic properties like binding energy from a bulk acceptor but also in the spatial extension of its wave function. Additionally, the cleavage surface which is needed in the STM experiment to access buried dopants introduces a symmetry reduction into the system that is not included in the bulk calculations. Thus, these calculations do not completely reproduce the observed asymmetric shape, and especially not the recently reported depth dependent changes of this asymmetry\cite{Mar2007b}.
In this paper we quantitatively study the evolution of the acceptor wave function with respect to its dependence on the interaction strength with surface-related and tip-induced effects by comparative images of acceptors in different depths under the surface. Surface strain fields and tip-induced electric fields are discussed on the basis of band structure calculations and can explain the symmetry reduction.

\section{Experimental Setup}
The experiments are performed in a low temperature STM operating in UHV at a base pressure better than $2\times10^{-11}$ mbar. Details of the experimental setup are given in ref.\cite{lot07prb}. The InAs samples are cleaved \textit{in situ} at room temperature and they are transferred to the precooled microscope where they reach the equilibrium temperature of 5.6~K within less than an hour after cleavage. The samples are conducting even at 4.2~K. The manganese doping concentration of $2\times10^{17}$cm$^{-3}$ establishes an impurity band with a few meV spectral width centered about 23 meV above the valence band edge\cite{Geo1989,Chi2004}.

\section{Results}

\begin{figure}
\includegraphics[scale=1.00]{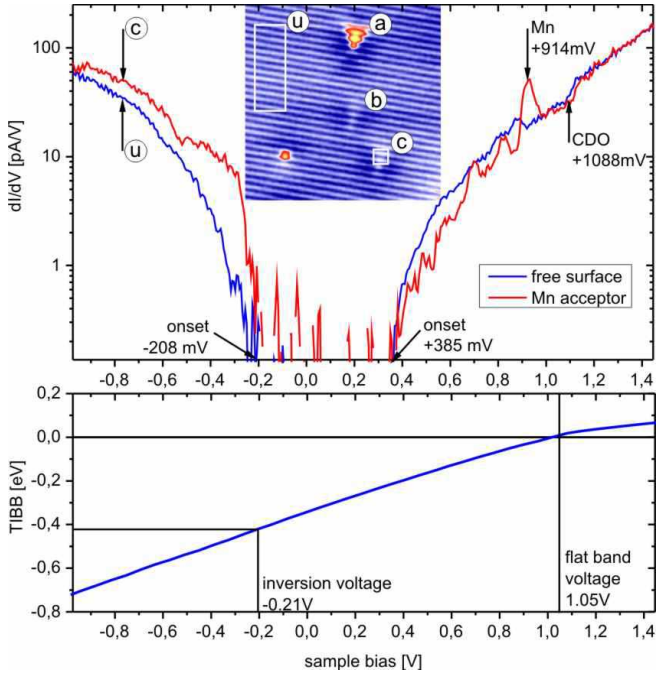}
%Man kann die Grafik auch noch beschneiden trim = 0 8 0 0,
\caption{\label{iumn} (Color online) Inset: (18$\times$18)nm$^2$ constant current topography of three subsurface Mn acceptors (a,b and c) under the InAs(110) surface. The topography is recorded at 1.0~V sample bias and 100~pA tunnel current. The Mn acceptors appear as asymmetric bow-tie like protrusions. Upper graph: local dI/dV characteristics acquired in the inset topography: The blue curve (labeled with u) corresponds to the undisturbed surface and the red one (labeled with c) was recorded directly above the lower Mn acceptor (c). Topographic setpoint for the I(V)-measurements is 2.0~V and 0.3~nA. At this setpoint the Mn acceptors have no impact on the topography. lower graph: numerically derived TIBB(V) dependence adjusted to the presented I(V)-spectroscopy. The characteristic bias voltages, inversion limit and flat band bias, are marked.}
\end{figure}

\begin{figure}
\includegraphics[scale=1.00]{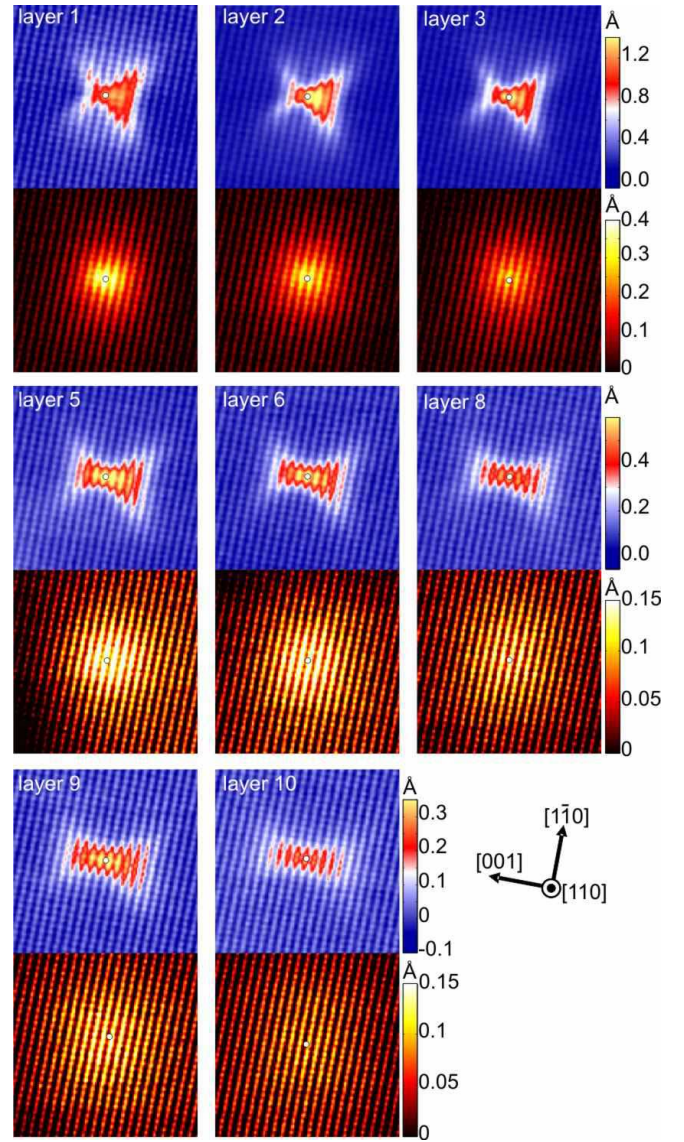}
%Man kann die Grafik auch noch beschneiden trim = 0 8 0 0,
\caption{\label{mb} (Color online) $(13\times13)$nm$^2$ zooms into the multibias topography. Each image doublet shows one acceptor. The blue-red colored image is recorded at +1.0V and the black-yellow colored images are acquired at -1.0V. The images of one row have the same color scaling. The adjacent color bar indicates the height scale for each row in \AA. The white circles show the location of the dopant atom under the surface as determined by the center-of-mass and contrast maximum of the circular contrast at -1.0V bias. }
\end{figure}

The first step of the analysis is the identification of the sample bias voltage at which the acceptor bound hole is imaged. The tip induced band bending (TIBB) present at the \{110\} surfaces of InAs has to be considered. It causes a non-trivial relation of sample energy scale and applied sample bias\cite{fee87,fee03}.
The constant current topography in Figure~\ref{iumn} presents an (18 $\times$ 18)nm$^2$  image of an atomically flat InAs(110) surface recorded at 1.0~V sample bias. The anisotropic bow-tie like contrasts of three subsurface Mn acceptors are visible. Two local differential conductivity (dI/dV) curves are acquired in this region (upper graph in Fig.~\ref{iumn}):
The blue one (denoted with u) is acquired in the white rectangle in the upper left corner of the topography and shows the dI/dV-signal of the undisturbed surface. The red spectrum (c) is recorded above the Mn contrast labeled with (c).
The lower graph in Fig.~\ref{iumn} presents the numerically derived TIBB(V)-dependence, which has been validated with spatially resolved I(V)-spectroscopy in the same manner as described earlier for GaAs\cite{lot07prb}. The actual TIBB(V) dependence is strongly affected by parameters that vary for different tips and thus need to be checked for each STM measurement. In order to identify the acceptor state in the I(V)-spectra, knowledge of the flat band bias voltage (TIBB=0~meV) is crucial. Thus, the tip work function, which determines the flat band bias, is experimentally evaluated. The TIBB(V) is calculated using this value (4.25~eV for the presented STS-measurement), a typical tip geometry (15nm tip apex radius, $90^\circ$ shank slope) and an estimated vacuum gap of 8~\AA (The numerical model introduced by Feenstra is used\cite{fee03}.). The calculated flat band bias is at 1.05~V sample bias. Delocalized charge density oscillations (CDO) appear as a conductivity step in both dI/dV-curves at 1088mV. This observation fixes the flat band bias to a slightly lower value which is in good agreement with the calculation.
The detection of the acceptor state is expected below the flat band bias. The prominent conductivity peak at +914mV that is solely observed above the acceptor contrast is therefore identified as the additional tunnel channel into the acceptor ground state. At positive bias it becomes accessible when the acceptor state is lifted above the Fermi energy. Thus, the spatial distribution of this conductivity peak is associated with the direct detection of the wave function of the acceptor bound hole. The comparison of the dI/dV-curves above and below this peak adds further confidence to this: for bias voltages lower than +0.9V the the acceptor state is below the Fermi energy, i.e., the acceptor is in its ionized state. No hole is bound to it and the dopant core's negative charge locally shifts the conduction and valence band upwards. The dI/dV-curve near the acceptor differs largely from the one recorded above the undisturbed surface. For bias voltages exceeding +0.9V the dopant core's negative charge is compensated by the bound hole. The surrounding area is no longer electrostatically distorted and both dI/dV-curves match.
Concluding this paragraph, the STM maps the probability density distribution of the Mn acceptor wave function for $>$+0.9V bias voltage.

The further analysis is done by topographic measurements. A total of 29~acceptor contrasts is acquired in a single atomically resolved (210 $\times$ 210)~nm$^2$ multibias measurement. Each line of the image is scanned subsequently with two different bias voltages before the scanner moves to the next line. Because two topographies are acquired quasi-simultaneously, thermal drift and piezo creep between them is negligible\cite{Gar2007}, and absolute positions in both images match to an accuracy better than one surface lattice constant. This was verified by comparing the positions of uncharged surface point defects between both images. Figure~\ref{mb} presents zooms in on eight different acceptor contrasts. The two biases are chosen such that the acceptor state is imaged in one topography while the acceptor core position can be determined in the other.
The first topography is recorded at +1.0 V, i.e., directly above the acceptor state's conductivity peak. The respective zooms are the blue-red colored images (upper image for each acceptor). According to the dI/dV-curves they are an image of the acceptor bound hole's spatial distribution.
The second topography is recorded at $-1.0$~V. At this bias the tunnel current is dominated by the valence band states and the acceptors exhibit circular symmetric protrusions. The negative acceptor charge has a circular symmetric Coulomb potential that influences the bands\cite{zhe94,kor01}. The center-of-mass and maximum of this contrast resembles the projected position of the acceptor core under the surface. It is indicated by white circles in Fig.~\ref{mb}.

The depth of each acceptor atom under the surface is determined as follows:
all visible acceptors are ordered to increasing depth under the assumption that the circular protrusion in the filled states image at -1.0V is strongest for the acceptor nearest to the surface and becomes fainter for deeper acceptors.
The acceptor contrasts in Fig.~\ref{mb} are ordered to increasing depth from top-left to bottom-right. The black-yellow colored images (lower image for each acceptor contrast) show the evolution of the circular contrast. To pinpoint not only the depth ordering but also the precise dopant depth, additional information is used: The symmetry center of the acceptor contrasts has to follow a certain ordering with respect to the host lattice\cite{saut04,dep01}.
Mn is a substitutional acceptor on the In site. The dominant empty states resonance at +1.0 V has its corrugation maxima above the In sites of the surface zig zag row\cite{kli03}. Therefore, an acceptor contrast in the first surface layer is centered directly on the corrugation maximum. If the acceptor is positioned in the second monolayer, the acceptor atom is located between the corrugation maxima. The acceptor contrasts in Fig.~\ref{mb} follow the alternating on-maximum, off-maximum ordering. Recent reports suggest that acceptors located in the two monolayers that form the surface have a different appearance\cite{Mar2007b,Kit2006}.
Thus, the label 'layer 1' in Fig.~\ref{mb} refers to the first subsurface layer. The acceptor core positions are determined for acceptors down to the 10th subsurface monolayer. The analysis shows that no acceptor was found in the 4th and 7th layer under the surface. About 4-5 additional depths were detected but the exact position of the respective acceptors could not be determined accurately any more due to the vanishing feature height in the filled states image. Nevertheless, it is worth noting that the STM could resolve acceptors that were up to 3~nm below the sample surface.
The ordered image sequence of the anisotropic acceptor contrasts (blue-red colored images in Fig.~\ref{mb}) shows a gradual shift from nearly triangular to nearly rectangular shapes. The acceptor in the first subsurface layer for example is of nearly triangular shape. The contrast maximum is shifted to the [00$\bar{1}$] side of the acceptor atom and the [001] side consists only of faint branches. An acceptor in the 10th subsurface layer on the other side appears as a nearly rectangular feature centered around the dopant site. Acceptors in intermediate depths exhibit an intermediate contrast.

\begin{figure}
\includegraphics[scale=1.00]{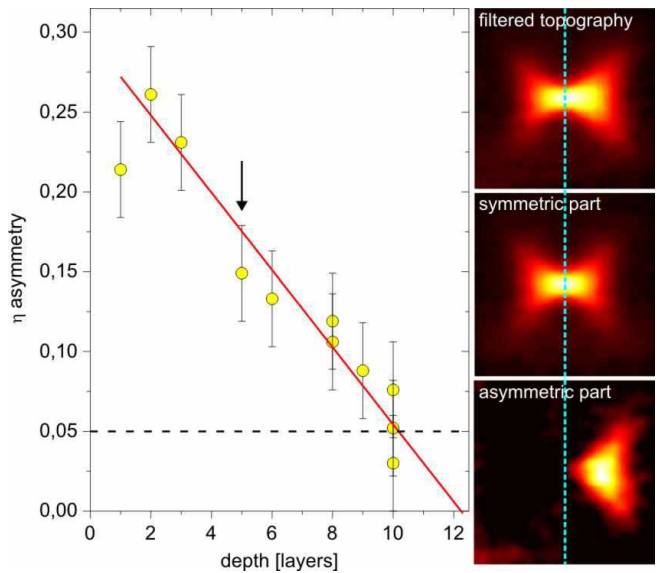}
%Man kann die Grafik auch noch beschneiden trim = 0 8 0 0,
\caption{\label{asym} (Color online) Asymmetry factor $\eta$ of the Mn acceptor contrasts plotted against dopant depth. These are the results of the symmetry analysis. If the acceptor in the first monolayer is excluded, the asymmetry decreases linearly to 0 with increasing depth. The lower limit to which the asymmetry can be detected is 0.05 which is reached for an acceptor in the 10th subsurface layer. The images at the right side demonstrate the symmetry analysis for an acceptor in the 5th layer (indicated by an arrow in the graph).}
\end{figure}

To gain a quantitative measure for this behavior, the degree of asymmetry with respect to the [001] direction is evaluated by image analysis as shown in Fig~\ref{asym}. The topography of each acceptor is decomposed into a symmetric and an asymmetric part with reference to the (001) mirror plane.
At first, the atomic corrugation of the surface states is suppressed in the images by FFT filtering to minimize the background signal (upper image of Fig.~\ref{asym}). The symmetric part $z_s(x,y)$ is evaluated with respect to a (001) mirror plane that cuts through the acceptor atom (middle image of Fig.~\ref{asym}). The symmetric part is subtracted from the topography, which results in an image of the asymmetric part $z_a(x,y)$ (lower image of Fig.~\ref{asym}). The degree of asymmetry of each acceptor contrast is described by the relative weight of the derived symmetric and asymmetric part. The quotient $\eta$ is a quotient of the integrals of the height information of symmetric and asymmetric image.

$$\eta={{\int{z_a(x,y)ds}}\over{\int{z_a(x,y)ds}+\int{z_s(x,y)ds}}}$$

It describes the ratio of asymmetric to symmetric components of the topography. The graph in Fig.~\ref{asym} plots $\eta$ for all acceptors of Fig.~\ref{mb} against the acceptor depth. The degree of asymmetry is as high as 27\% for the acceptor in the 2nd layer. With increasing depth the asymmetry decreases nearly linearly. The slope of the linear fit gives a decrease of 0.024 per monolayer depth. Acceptors in the 10th subsurface layer are fully symmetric within the accuracy of this analysis. The detection limit is estimated by performing the same analysis with the mirror plane (1$\overline{1}$0).  The Mn acceptor is mirror symmetric with that plane, but the value $\eta$ varies between 0 and 0.05 for this direction because of residual noise (this limit is indicated by a dashed line in Fig.~\ref{asym}). The uncertainty of the symmetry analysis is lower than the measurement noise, it is $\Delta\eta= 0.03$. If the linear fit is extrapolated to deeper acceptors, the 12th layer acceptor would be completely symmetric. As a result acceptors buried below the 10th to 12th layer under the surface appear as rectangular contrasts that are mirror symmetric with respect to both the (1$\overline{1}$0)  plane and the (001) plane. Acceptors located within the first 10 subsurface layers have a (001) mirror asymmetry. The [00$\overline{1}$]  side of the acceptor contrast is more pronounced than the [001] side.

\section{Discussion}
\begin{figure}
\includegraphics[scale=1.0]{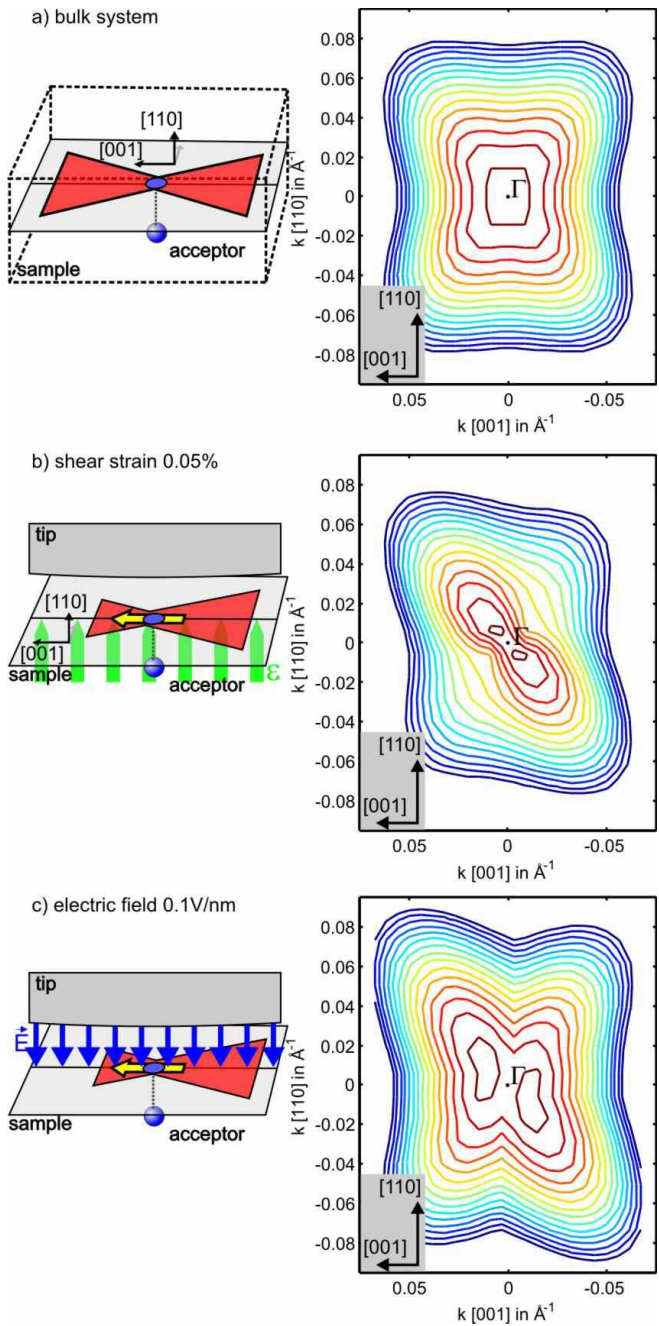}
%Man kann die Grafik auch noch beschneiden trim = 0 8 0 0,
\caption{\label{bands} (Color online) Band structure of InAs in a plane defined by [001] and [110]. The colored lines are iso-energy lines of the upmost valence band. The plots show about 10\% of the Brillouin zone. The calculated situations are sketched to the left of each plot. (a) No additional symmetry reducing field is applied, i.e., bare InAs band structure. This results in a symmetric contrast with reference to [001]. (b) A shear strain is applied that accounts for the strain induced by the surface relaxation. The In atoms are displaced by 0.05\% of the (110) layer distance. (c) An electric field is applied along the [110] direction with a strength of 0.1 V/nm. Both (b) and (c) induce a (001) mirror \underline{a}symmetry in the contrasts at the [110] surface.}
\end{figure}

On the basis of the local I(V)-spectroscopy (Fig.~\ref{iumn}) we conclude that the asymmetric contrasts at +1.0V are an image of the acceptor bound hole, i.e., they resemble the probability distribution of the acceptor wave function at the surface. And indeed, the observed probability density distribution of deeply buried acceptors has a nearly rectangular shape. This fits well with the theoretical expectation for a bulk acceptor as calculated for example by effective mass \cite{Mon2006} or tight-binding methods\cite{fla05, Mar2007b}.
The depth dependent measurements demonstrate that the probability density distribution for acceptors near to the surface are deformed compared to the deeply buried acceptors.

The presented system offers the unique possibility to study the impact of the surface's symmetry reduction effect on a dopant's wave function. Certainly, the spatial extension of the acceptor wave function is decreased by the vertical confinement of the surface. Additionally, the half-space geometry (1/2 semiconductor and 1/2 vacuum) will affect other properties such as the binding energy, as well (see e.g. refs.\cite{Krc2000, tei2007b}). Unfortunately, a realistic description of the surface in sophisticated quantum mechanical models, e.g., tight-binding calculations or ab-initio density functional theory approaches, exceeds today's computational abilities.

However, the symmetry properties of the acceptor wave function may be elucidated by considerations based on the bulk band structure: the acceptor state is a hybrid of the upper valence band states. The energy window of the valence band needed to form the localized state approximately equals the binding energy of this state \cite{sto1975}. The Mn acceptor in InAs is 23~meV above the valence band maximum \cite{Chi2004}, so about 10\% of the Brillouin zone participates in the hybridization. The symmetry of its wave function is determined by the host crystal's band structure. If the band structure is symmetric along a certain direction, the wave function will be symmetric, as well. Anisotropies in the wave function can only develop when the band structure exhibits this asymmetry. The bulk bands are to good approximation cubic in this range. In particular, they are symmetric with respect to the [001] direction. Effects that break this symmetry are known but usually considered to be small in the bulk semiconductor. The so-called k-linear terms for example cause a splitting of less than 1~meV at the valence band edge in the bulk\cite{Car1988}.
Figure~\ref{bands}a presents a band structure calculation for bulk InAs. Empirical pseudo-potentials\cite{har00,che79} were used and the spin-orbit interaction (SOI) was explicitly included\cite{gre1997}. The band structure is evaluated for a cut defined by a plane consisting of the [001] and [110] directions.  This cut visualizes the symmetry properties of the InAs band structure that will affect the shape of acceptor contrasts at the (110) surface relative to the [001] direction. The graph in Fig.~\ref{bands} shows energy contour lines of the highest valence band.  The plotted section has an extension of about 10\% of the Brillouin zone.  For the bulk system without any symmetry reducing field (Fig.~\ref{bands}a) the well-known shape (symmetry c$_{2v}$ ) is reproduced. The band is symmetric with reference to the (001) mirror plane, i.e., the [00$\overline{1}$] and the [001] part of the graph are identical. The resulting acceptor wave function will inherit this symmetry and the acceptor contrast at the surface is symmetric as depicted in the sketch (Fig.~\ref{bands}a, left). This agrees well with the previously reported theoretical predictions for the Mn acceptor wave function\cite{Mon2006,yak04} and the measured contrast of acceptors located in the 10th subsurface layer (refer to Fig.~\ref{mb}, layer 10).
In contrast to that, our STM analysis shows that acceptors close to the surface exhibit a strong asymmetry along [001] (refer to Fig.~\ref{mb} \& Fig.~\ref{asym}) while they remain symmetric with reference to the [$\overline{1}$10] mirror plane. This asymmetry cannot be described with the bulk band structure only. Obviously, the cleavage surface induces a symmetry breaking that lifts the cubic symmetry with reference to the [001] mirror plane but preserves it along the perpendicular [$\overline{1}$10] direction. In the following, two effects that introduce strong changes to the band structure will be discussed: Local strain fields and strong electric fields. Both are present under the STM tip at the relaxed surface.

The atoms in the first few layers of the InAs(110) surface relax outwards. The relaxation is usually treated in self-consistent pseudo-potential calculations\cite{Alv1991, che79, zun80} and investigated by low energy electron diffraction (LEED)\cite{kah1978, ton1978}. In terms of strain this outward relaxation decomposes into a hydrostatic component and uniaxial component. The uniaxial strain is along [110] for the relaxation. Neither the hydrostatic nor the uniaxial components induce symmetry breaking with respect to [001]. However, the Zincblende crystal has the additional property that uniaxial strain along the $\langle110\rangle$ directions results in shear components in the strain tensor \cite{bir1974}. They correspond to the off-diagonal components $\varepsilon_{ij}$ of the strain tensor, while the uniaxial strain is part of the diagonal components $\varepsilon_{ii}$. In recent experiments it was shown that even small [110] uniaxial strain (that can be applied by an external vice) leads to large anisotropies in the electron propagation properties in GaAs \cite{cro2005a}. This gives rise to the idea that the [110] uniaxial strain induces the observed symmetry reduction along [001].
As a first approximation to the surface relaxation the induced shear is modeled by a slight shift of the In sublattice into the [110] direction.
The influence of the symmetry reducing strain field on the highest valence band is shown in Fig.~\ref{bands}b. The results show a prominent symmetry reduction with reference to the (001) mirror plane. Already very small distortions induce a considerable asymmetry. The graph Fig.~\ref{bands}b shows the valence band for an In displacement of 0.05~\% of the [110] monolayer distance. The valence band becomes elongated along [111] and compressed along [11$\overline{1}$]. An acceptor wave function in this environment will extend further along [11$\overline{1}$] than along [111]. Thus, the probability density on the [110] surface will extend further along [00$\overline{1}$] than it does along [001]. The resulting contrast properties are depicted in the sketch of Fig.~\ref{bands}b. This matches with the measurement: the asymmetric bow-tie like contrasts are more pronounced on the [00$\overline{1}$] side of the dopant atom. These findings are corroborated by a recent report on Mn acceptors located in the strain field of a quantum dot. The in-plane strain has a strong influence on the wave function shape of a dopant and distorts the acceptor contrast into the direction of the quantum dot\cite{yak2007}.

The second effect that is capable of producing the [001] asymmetry of the Mn contrast is the tip-induced electric field. The tip exhibits an electric field penetrating into the semiconductor. It is parallel to the surface normal of the cleavage plane [110] due to the STM geometry. Typical field strengths are on the order of $10^5$~V/cm to $10^6$~V/cm. The STM images of the acceptor wave function show that the relative weight of the acceptor wave function shifts perpendicular to this electric field. It is worth noting that an electrostatic distortion of the wave function due to such an electric field, like e.g., the Stark effect, would only produce changes that are symmetric with reference to the (001) mirror plane. An elongation or compression of the wave function along [110] would not explain the observed asymmetry. An effect is needed, that acts differently for the [001] and [00$\overline{1}$] wave vector components. The spin-orbit interaction (SOI) provides this kind of symmetry reduction in the band structure\cite{Car1988}. The above calculation is extended to implement a homogenous electric field in the [110] direction. It is introduced to the Hamiltonian via the SOI term in the form of the Rashba Hamiltonian\cite{yu1996}.
As illustrated by the sketch in Fig.~\ref{bands}c the electric field resembles a structure inversion asymmetry (SIA)\cite{winr2003}. Its effect on the highest valence band is shown in the graph of Fig.~\ref{bands}c for an electric field of 0.1~V/nm. The valence band distortion is similar to the previously discusses strain field. The valence band is elongated along the [111] direction and compressed along the [11$\overline{1}$] direction. Thus, the effect on the acceptor contrast at the [110] surface will be comparable. The [00$\overline{1}$] side of the bow-tie like contrast will be more pronounced than the opposite side.
The calculated deformation of the valence band is caused by the combination of the bulk inversion asymmetry (BIA) of the Zincblende crystal and the external field induced SIA. The BIA preserves the cubic shape of the valence bands, i.e. their elongation in all $\langle111\rangle$ directions and compression in the $\langle100\rangle$ directions. The spin splitting of the bands due to SIA has a different dependence. The sign of the spin splitting due to SIA and BIA in the [111] direction is the same. Both effects add up. In the perpendicular [11$\overline{1}$] direction SIA and BIA have opposite sign and decrease each other. The sum of both effects induces the symmetry reduction over [001].

The valence band shape under influence of strain and electric field is similar. Both effects act similarly on the host crystal's band structure and will thus introduce similar asymmetries in the acceptor state's wave function. The calculated valence bands look alike.
Comparison of Fig.~\ref{bands}b and Fig.~\ref{bands}c yields estimated relative strengths of both effects:
Already small shear strain of 0.05\% ( [110] monolayer distance) is capable of producing considerable mirror asymmetry with reference to (001). On the other side a strong electric field in [110] direction of 0.1 V/nm (=$10^6$V/cm) is used to introduce a similar effect solely by electric fields. Typical tip-induced electric fields on III-V semiconductors doped in the order of $10^{17}cm^{-3}$ may reach this order of magnitude. Thus, both effects strain and electric field are capable of producing the observed asymmetry are not distinguishable by the presented measurements. A definitive experimental differentiation remains to be shown.

In summary the symmetry considerations of the band structure demonstrate that the strain field of the surface relaxation and the tip-induced electric field reduce the symmetry of the bulk band structure. The valence bands gradually develop a (001) mirror asymmetry for both cases. This is the same symmetry reduction as observed in the experiment. The calculations indicate that the acceptor contrast should be more pronounced on the [00$\overline{1}$] side for the [110] cleavage surface, which is supported by the measurement (compare sketches in Fig.~\ref{bands}b,c with STM topographies in Fig.~\ref{mb}).
Additionally, both symmetry reduction effects are limited to a narrow layer at the surface (compare with Fig.~\ref{asym}): Surface relaxation calculations indicate that the strain field rapidly decays for increasing depth under the surface\cite{Alv1991,kah1978}. The tip-induced electric field decreases linearly into the bulk sample because the tip-induced space-charge layer is of parabolic shape\cite{fee03}. This complies with the observation that deeper buried acceptors are more symmetric.

\section{Conclusion}
Mn acceptors in InAs are analyzed with high resolution multibias topographic measurements.
The anisotropic bow-tie like features at subsurface acceptors are identified as an image of the probability density distribution of the acceptor state. This is validated by local I(V)-spectroscopy. Comparative topographic measurements reveal a monotonous evolution from a nearly triangular shaped contrast to a rectangular one with increasing distance of the dopant atom from the (110) cleavage surface. Acceptors located within the first ten subsurface layers of the semiconductor show an asymmetry with reference to the (001) mirror plane. The degree of asymmetry can be as high as 27\%. Symmetry reduction effects at the surface like strain originating from the surface relaxation and electric fields induced by the STM tip are discussed as source of the observed asymmetry. The measured contrasts for acceptors buried below the 10th subsurface layer are in good agreement with theoretical predictions for the probability density distribution of the bulk acceptor wave function. Thus, the (001) asymmetry measured for most acceptor contrasts is traced back to the interplay of symmetry reduction effects at the surface and the cubic bulk band structure of the host crystal. These findings demonstrate that impurities in different depths under the surface give access to the evolution of the acceptor wave function in environments with varying anisotropy and/or reduced dimensionality.

\begin{acknowledgments}
We thank J. Wiebe, F. Marczinowski and P. M. Koenraad for valuable discussions.
This work was supported by DFG-SFB 602 - \emph{Complex structures in condensed matter from atomic to mesoscopic scales}, DFG-SPP 1285 - \emph{Semiconductor Spintronics} and the German National Academic Foundation.
\end{acknowledgments}

\end{document}